\documentclass[lettersize,journal]{IEEEtran}

\usepackage[T1]{fontenc}
\usepackage[caption=false,font=normalsize,labelfont=sf,textfont=sf]{subfig}
\usepackage{cite}
\usepackage{booktabs}
\usepackage[noend]{algpseudocode}
\usepackage{algorithmicx}
\usepackage[ruled]{algorithm2e}
\usepackage{amsmath}
\usepackage{amssymb}
\usepackage{multirow}
\usepackage{tabularx}
\usepackage[table]{xcolor}
\usepackage{url}
\usepackage{colortbl}
\usepackage{exscale}
\usepackage{relsize}
\usepackage{colortbl}
\usepackage{textcomp}
\usepackage{stfloats}
\usepackage{url}
\usepackage{verbatim}
\usepackage{graphicx}
\usepackage{bm}

\definecolor{mygray}{gray}{.9}
\definecolor{mybrown}{gray}{.8}
\definecolor{mycolor}{gray}{.7}
\usepackage{graphicx}
\graphicspath{{figures/}}
\usepackage{threeparttable}

\makeatletter
\newcommand*\bigcdot{\mathpalette\bigcdot@{.5}}
\newcommand*\bigcdot@[2]{\mathbin{\vcenter{\hbox{\scalebox{#2}{$\m@th#1\bullet$}}}}}
\makeatother
\usepackage{amsmath}
\usepackage[cmintegrals]{newtxmath}

\begin{document}

\title{Multi-Condition Fault Diagnosis of Dynamic Systems: A Survey, Insights, and Prospects}

\author{Pengyu Han, Zeyi Liu, Xiao He,~\IEEEmembership{Senior Member,~IEEE}, Steven X. Ding, and Donghua Zhou,~\IEEEmembership{Fellow,~IEEE}
	\thanks{This work was supported by National Natural Science Foundation of China under grant 62473223 and 624B2087, and Beijing Natural Science Foundation under grant L241016. (\emph{Corresponding author: Xiao He.})
		
		Pengyu Han, Zeyi Liu and Xiao He are with the Department of Automation, Tsinghua University, Beijing 100084, P. R. China. (emails: hpy24@mails.tsinghua.edu.cn, liuzy21@mails.tsinghua.edu.cn, hexiao@tsinghua.edu.cn).
		
		Steven X. Ding is with the Institute for Automatic Control and Complex Systems, University of Duisburg-Essen, 47057 Duisburg, Germany. (e-mail: steven.ding@uni-due.de).
		
		Donghua Zhou is with the College of Electrical Engineering and Automation, Shandong University of Science and Technology, Qingdao 266000, P. R. China, and also with the Department of Automation, Tsinghua University, Beijing 100084, P. R. China. (e-mail: zdh@mail.tsinghua.edu.cn).
	}
}

\maketitle
\begin{abstract}
With the increasing complexity of industrial production systems, accurate fault diagnosis is essential to ensure safe and efficient system operation.
However, due to changes in production demands, dynamic process adjustments, and complex external environmental disturbances, multiple operating conditions frequently arise during production. 
The multi-condition characteristics pose significant challenges to traditional fault diagnosis methods. 
In this context, multi-condition fault diagnosis has gradually become a key area of research, attracting extensive attention from both academia and industry.
This paper aims to provide a systematic and comprehensive review of existing research in the field. 
Firstly, the mathematical definition of the problem is presented, followed by an overview of the current research status. 
Subsequently, the existing literature is reviewed and categorized from the perspectives of single-model and multi-model approaches. 
In addition, standard evaluation metrics and typical real-world application scenarios are summarized and analyzed. 
Finally, the key challenges and prospects in the field are thoroughly discussed.
	
\end{abstract}

\def\abstractname{Note to Practitioners}
\begin{abstract}
	As industrial production systems grow more complex, traditional fault diagnosis methods face considerable challenges in addressing multi-condition issues arising from changing production demands, dynamic process adjustments, and disturbances from complex external environments.
	As a result, multi-condition fault diagnosis has become a key area of research and has garnered extensive attention in recent years.
	This paper presents a systematic review of multi-condition fault diagnosis for dynamic systems, ranging from developmental overview to key frameworks, evaluation metrics, and typical application scenarios.
	To assist researchers in quickly grasping the current trends in the industry, this paper further summarizes the challenges and future prospects within the field.
	
\end{abstract}



\section{Introduction}
\IEEEPARstart{W}{ith} the continuous advancement of automation, the scale and complexity of industrial production systems are also steadily increasing \cite{Chen2023d, liu2023discrimination, he2023real,liu2024chinesedynamic}. 
Modern industrial systems typically consist of multiple interconnected subsystems, operating in highly automated environments that impose stricter demands on system stability and reliability \cite{ma2023adaptive, geng2024reliable}. 
A failure in such systems can significantly decrease production efficiency, disrupt regular production schedules, and even jeopardize operational safety, leading to substantial economic losses or even casualties \cite{Zhang2021, Hei2024, Su2021, liu2024online}.
Against this backdrop, the development of efficient and accurate fault diagnosis methods has become critically important. 
Accurate fault diagnosis not only facilitates the timely detection and localization of system anomalies but also prevents secondary hazards, thereby improving system reliability while reducing maintenance costs. 
As a result, fault diagnosis has increasingly emerged as a focal point of research interest in recent years \cite{han2024imbalanced, Huo2022, Jin2020, liu2024review}.

Fault diagnosis methods are generally categorized into model-based \cite{ ding2008model,zhong2023overview} and data-driven approaches \cite{Fan2024a}. 
Model-based methods rely on constructing accurate mathematical models of the system. By applying identical control signals to both the physical system and its mathematical model, residual signals are generated and subsequently analyzed using evaluation and classification functions to detect and classify faults \cite{zhao2022model}. 
Nevertheless, as the complexity of modern industrial equipment increases, it has become progressively challenging to develop precise mathematical models, thereby limiting the practical applicability of model-based approaches. 
In contrast, advancements in sensor technology have enabled industrial processes to acquire vast amounts of operational data, which have laid a strong foundation for the development of data-driven methods \cite{An2020, Zhang2022a, liu2022fault}. 
Data-driven approaches encompass techniques such as signal processing \cite{Xu2023a, Borghesani2013 } and machine learning \cite{Zhao2023, Lu2023a }. 
Notably, with the rapid development of artificial intelligence \cite{yue2023metal,dong2023neural}, machine learning-based fault diagnosis methods have been widely applied in complex systems. 
These methods demonstrate superior capabilities in feature extraction and learning, achieving remarkable performance in numerous fault diagnosis tasks \cite{Wu2024, Yang2023a,Yang2023, Li2022a}.

However, in many industrial systems, changes in production demands, dynamic adjustments in processes, or complex variations in external environments often lead to significant shifts in data distributions \cite{Fan2024, Zou2021, Liu2025, Yuyan2024}. 
Such distribution shifts may lead to diagnostic models trained under specific operating conditions failing to sustain stable performance in other conditions, or even result in complete failure. 
This issue is particularly pronounced in industrial scenarios characterized by diverse and frequently changing operating conditions, posing significant challenges to the adaptability, robustness, and generalization capabilities of fault diagnosis methods \cite{Zhu2024, Du2020, Li2024, Yuyan2024}. 
\emph{Multi-condition fault diagnosis} (MCFD) aims to address this issue, with the primary objective of ensuring that diagnostic models can deliver consistent and reliable performance across diverse operating conditions. 
Unlike single-condition diagnosis, MCFD must account for the substantial differences in data distributions across various conditions \cite{Zhang2020, Xu2021, Chen2021, Zhang2022}, as well as the unpredictability posed by unknown conditions \cite{Geng2020, Zhang2020a, Liu2022}. 
This introduces multiple challenges, including how to construct models that can accommodate the complex nonlinear relationships between operating conditions, how to address performance degradation caused by data distribution shifts, and how to achieve efficient and accurate fault identification under unknown conditions. 
To overcome these challenges, MCFD has emerged as a key research focus in the field of fault diagnosis in recent years, garnering significant attention from both academia and industry \cite{ Li2021, Tang2022}.

\begin{figure}[htpb]
	\centering
	\includegraphics[width=3.4in, keepaspectratio]{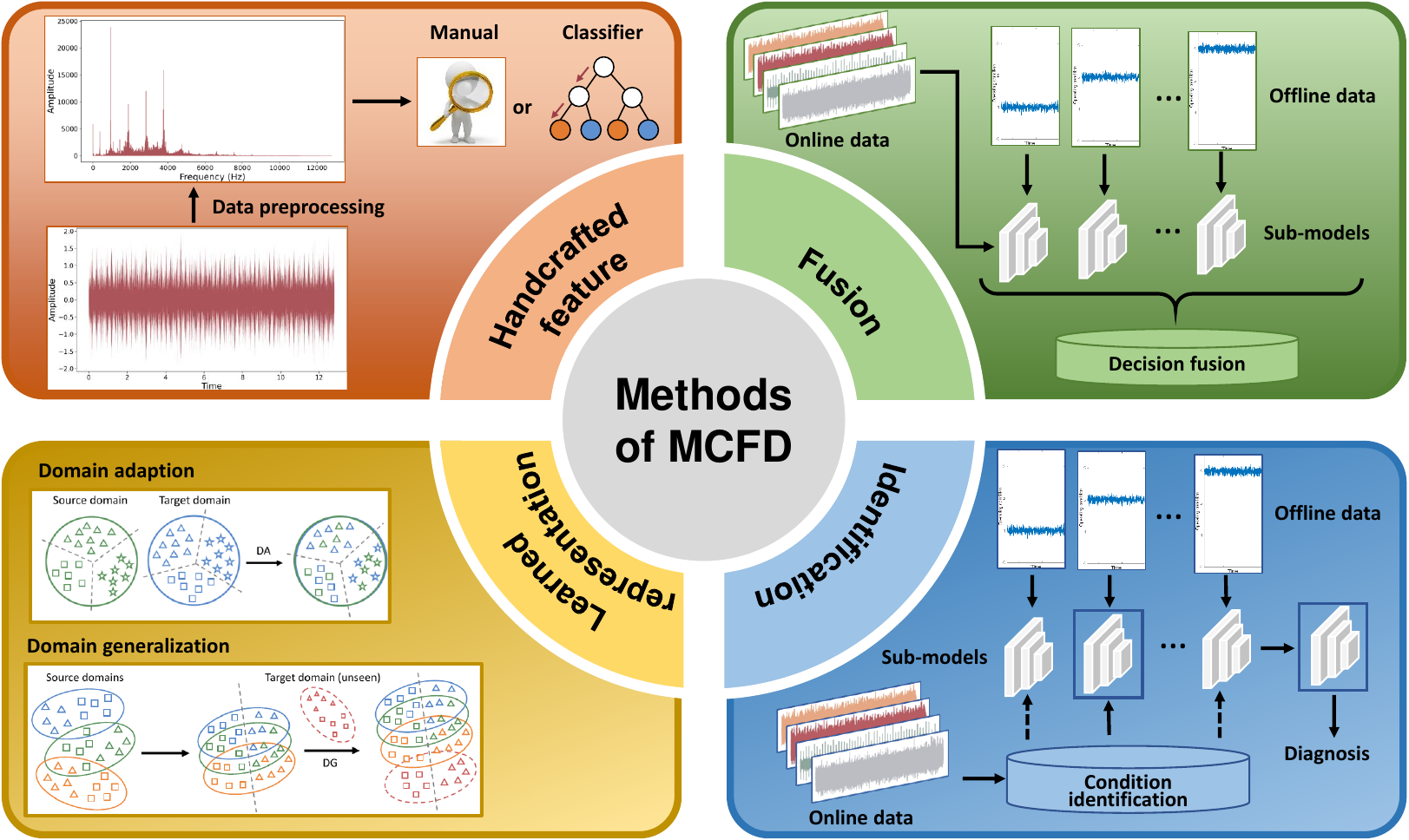}
	\caption{Overview of different methods in multi-condition fault diagnosis.}
	\label{TOC}
\end{figure}

This paper reviews approximately 200 relevant studies in the field of MCFD over the past two decades, aiming to provide a comprehensive and systematic overview of this area. 
The related materials organized for this review are available at \url{https://github.com/THUFDD/Multi-Condition-Fault-Diagnosis}, and the repository will be continuously updated.
In Section II, we present the mathematical definition of the MCFD problem and provide a brief overview of the current research status in the area. 
Sections III and IV focus on single-model and multi-model MCFD approaches, respectively, as shown in Fig. \ref{TOC}. These sections aim to offer an in-depth review of related research while analyzing the strengths and weaknesses of each method.
Section V outlines the commonly used evaluation metrics for MCFD tasks and discusses their distinct features.
Additionally, in Section VI, we explore typical application scenarios for MCFD, including mechanical systems and chemical processes. 
Finally, Section VII provides a summary of the paper and discusses future research directions and development prospects.

\section{Problem Statement}
\subsection{Problem Formulation}
\begin{figure}[!htbp]
	\centering
	\subfloat[]{
		\includegraphics[width=0.45\textwidth]{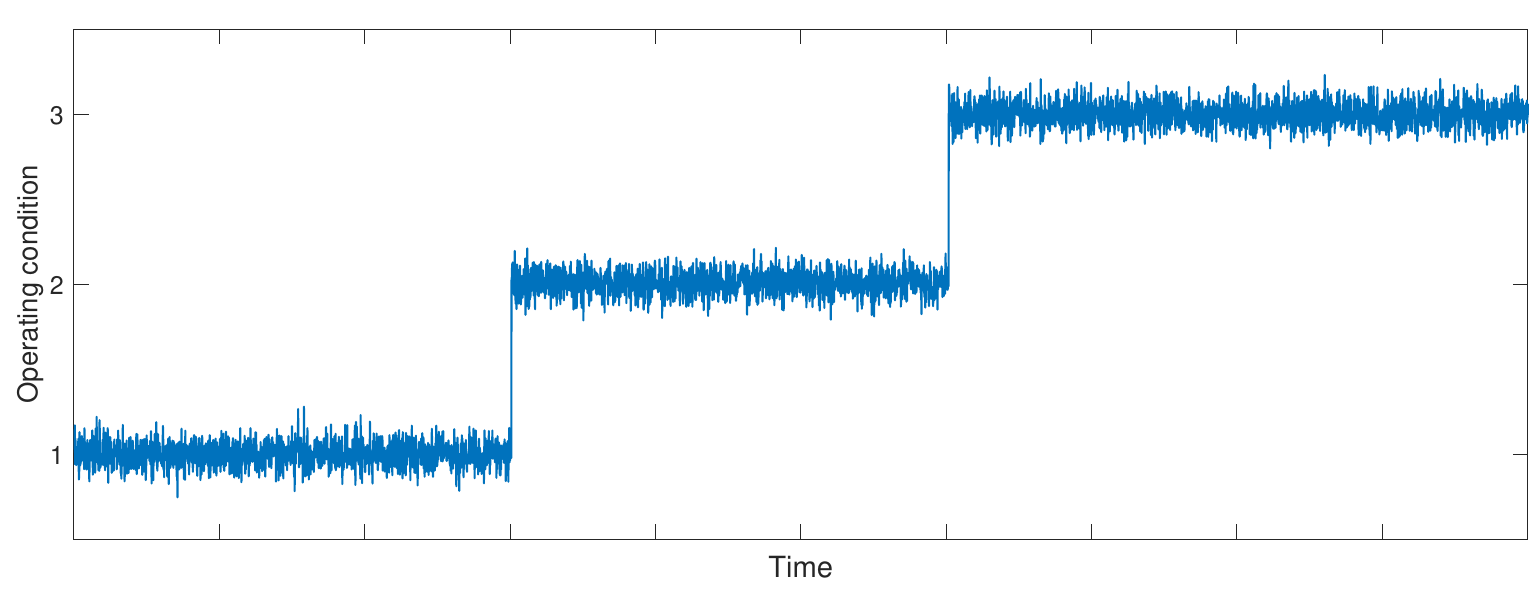} \label{steady}
	}\\
	\subfloat[]{
		\includegraphics[width=0.45\textwidth]{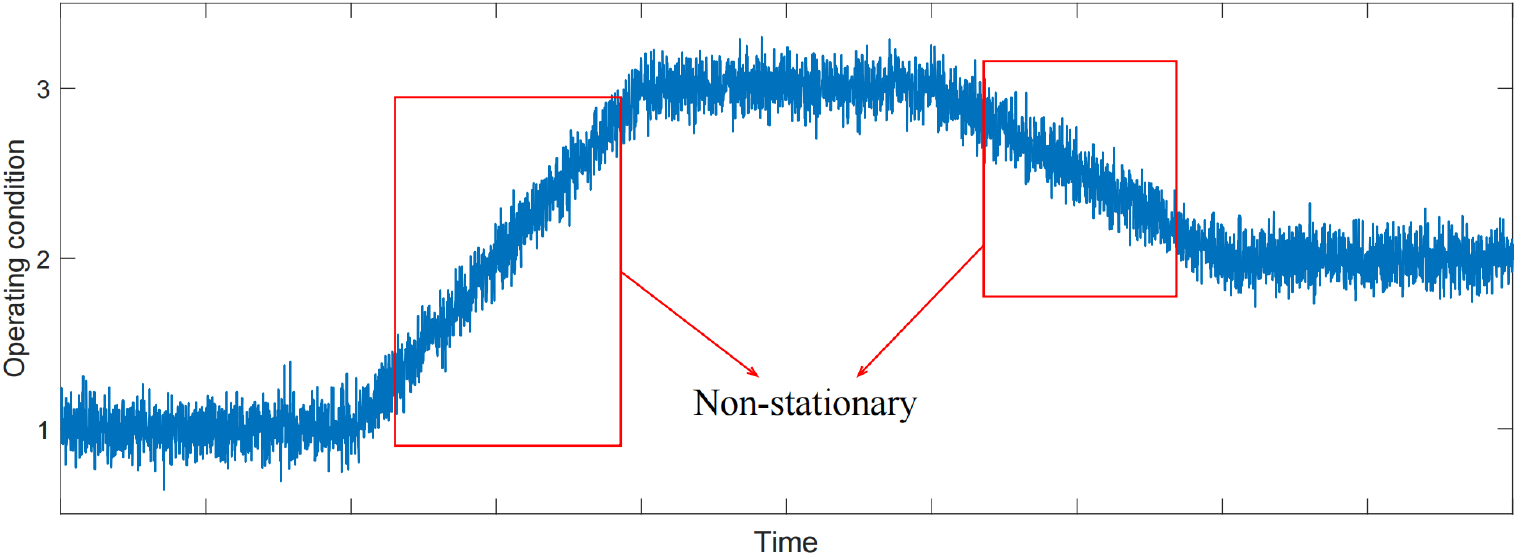} \label{unsteady}
	}
	\caption{Schematic diagram of different multi-condition scenarios. (a) Steady multi-condition scenario. (b) Unsteady multi-condition scenario.}
\end{figure}

Due to the influences of load fluctuations, external environmental disturbances, and other factors, industrial processes can be considered as stochastic processes \cite{Zhang2023multimode}, denoted as:
\begin{equation} 
	\{ X(\omega, t):t = 0, \pm 1, \pm 2, \dots \}, 
\end{equation}
where $\{ X_{t_1}, \dots, X_{t_n} \}$  represents the random variables in this process. The \emph{cumulative distribution function} (CDF) of this random variable is defined as follows:
\begin{equation} 
	F(x_{t_1}, x_{t_2}, \dots, x_{t_n}) = P(X_{t_1} \leq x_{t_1}, X_{t_2} \leq x_{t_2}, \dots, X_{t_n} \leq x_{t_n}).
\end{equation}
If for any time points $t_1,t_2,\dots,t_n$ and any time shift $\tau$, the following constraint is satisfied:
\begin{equation}
	F(x_{t_1}, x_{t_2}, \dots, x_{t_n}) = F(x_{t_1+\tau}, x_{t_2+\tau}, \dots, x_{t_n+\tau}),
\end{equation}
then the process is considered stationary, indicating the cumulative distribution function does not vary with time \cite{kan2015review}. If this constraint is not satisfied, the process is regarded as non-stationary. 
In the following description, if a multi-condition scenario consists solely of stationary processes, it is referred to as \emph{steady}. Conversely, if the multi-condition scenario includes at least one non-stationary process, it is termed \emph{unsteady}. Figs. \ref{steady} and \ref{unsteady} illustrate the schematic diagrams of steady and unsteady scenarios, respectively.

\subsection{Research status of MCFD}
\begin{figure}[!htbp]
	\centering
	\includegraphics[width=3.3in, keepaspectratio]{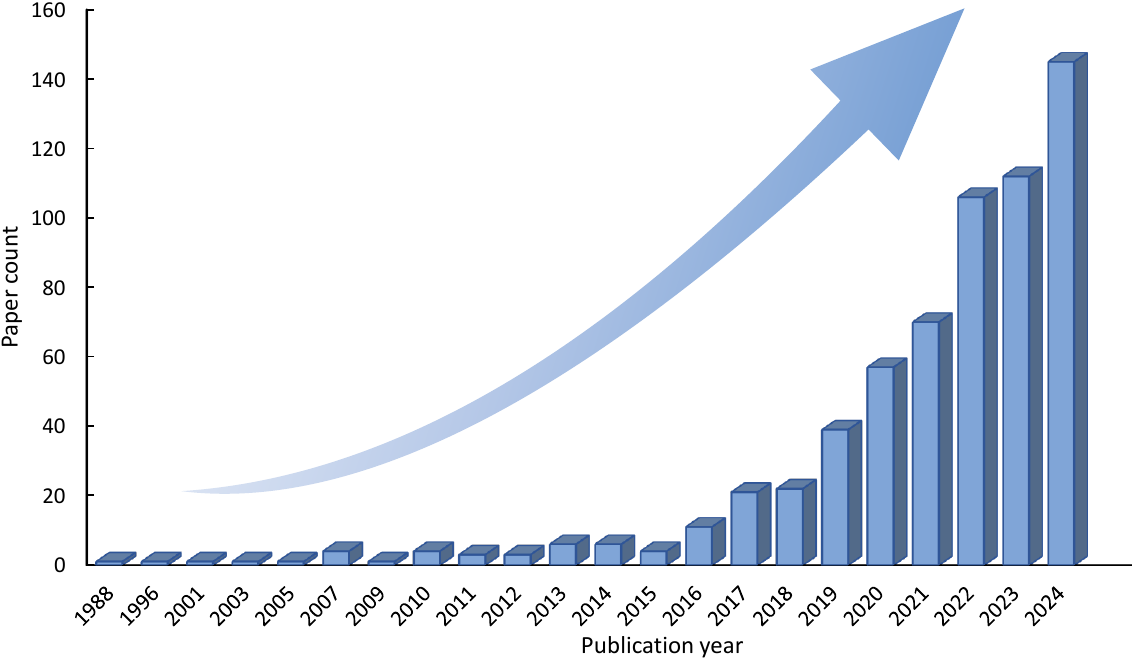}
	\caption{Statistical chart of publication counts for various methods in MCFD based on Web of Science data.}
	\label{paper_number}
\end{figure}

Data-driven fault diagnosis can be categorized into two approaches. The first approach is the \emph{fault detection and isolation} (FDI) method, which involves two sequential steps \cite{zhouFaultIsolationBased2016, wang2024fault}. Initially, fault detection is performed to identify anomalies in the process. If an anomaly is detected, the model subsequently triggers the fault isolation module to locate the fault source. The second approach considers the fault diagnosis problem as a fault classification task, directly predicting fault types using classification models.
In recent years, MCFD has garnered increasing attention. 
By performing keyword searches on Web of Science, it is observed that the number of publications in the field has exhibited a steady upward trend, as illustrated in Fig. \ref{paper_number}.

While considerable research has been conducted on MCFD, the field still lacks a systematic review that thoroughly synthesizes existing studies.
Therefore, this article aims to fill this gap by providing a comprehensive review of MCFD, highlighting key developments, methodologies, and future research opportunities.

\begin{figure}[!htpb]
	\centering
	\includegraphics[width=3.3in, keepaspectratio]{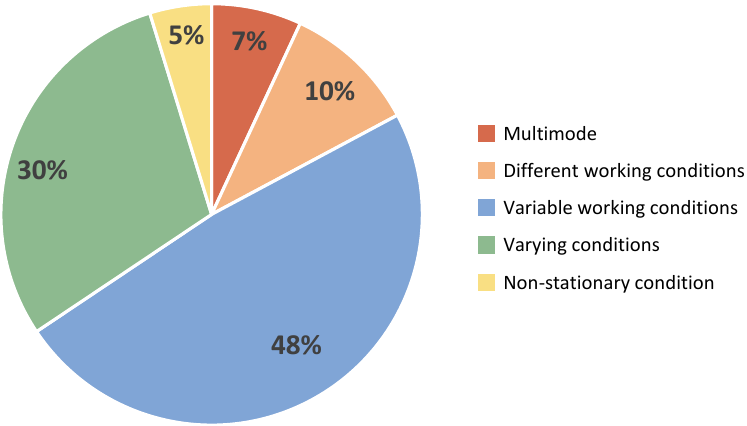}
	\caption{Proportion of different descriptions for MCFD in the literature.}
	\label{name_pie}
\end{figure}

In the existing literature, various terms have been used to describe MCFD, as shown in Fig. \ref{name_pie}, such as "multimode," "multiple operating modes," "variable working conditions," "time-varying operating condition," "different working conditions," etc. These terms convey the same concept. Therefore, to maintain consistency, the term "multiple operating conditions" or "multi-condition" will be adopted in subsequent discussions to describe variations such as changes in load, speed, or production schemes.

\subsection{Classification of existing MCFD approaches}
\begin{figure}[htpb]
	\centering
	\includegraphics[width=3.3in, keepaspectratio]{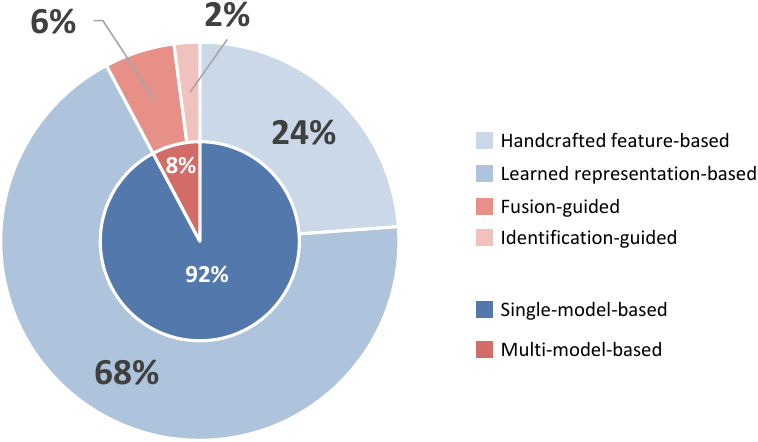}
	\caption{Proportion of different frameworks for MCFD methods in the surveyed literature.}
	\label{framework_pie}
\end{figure}

Based on a comprehensive literature review, current MCFD methods can be broadly categorized into two types: single-model-based approaches and multi-model-based approaches. 
Single-model approaches focus on utilizing a unified model to reduce the interference of condition-specific information with fault-specific information, enabling the extraction of fault-invariant features and ensuring satisfactory accuracy across various operating conditions. 
In contrast, multi-model approaches leverage offline multi-condition data to construct separate sub-models for each operating condition, incorporating condition information during fault diagnosis.
From the perspective of task applicability, single-model approaches are more suitable for tasks where the differences in data distribution across operating conditions are relatively small. These methods are particularly advantageous in scenarios involving diverse operating conditions or the presence of unknown conditions. In contrast, multi-model approaches are more appropriate for scenarios with a limited number of well-defined and known operating conditions. Their advantages become evident when there are substantial distributional differences between operating conditions and sufficient data is available for each condition.
Sections III and IV will explore these two categories of methods in detail. 
Fig. \ref{framework_pie} depicts the approximate proportions of various frameworks in the surveyed literature.

\section{Single-model-based MCFD methods}
Single-model methods can be divided into handcrafted feature-based and learned representation-based frameworks. 
The former offers the advantage of intuitive and partially interpretable feature extraction processes but requires additional classifiers to produce the final diagnostic results.
In comparison, the latter typically employs end-to-end models, which can directly generate diagnostic outcomes in a single step.
However, as learned representation-based frameworks rely on machine learning for feature representation learning, they often exhibit a "black-box" nature, resulting in weaker interpretability.
In this section, a detailed review of the existing methods based on these two types of frameworks will be provided.

\subsection{Handcrafted feature-based framework}
\begin{figure}[!htpb]
	\centering
	\includegraphics[width=2.2in, keepaspectratio]{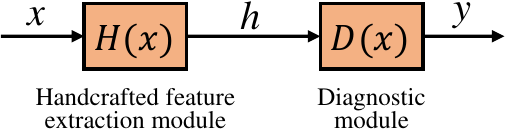}
	\caption{Structural schematic of the handcrafted feature-based framework.}
	\label{handcrafted}
\end{figure}

In the field of multi-condition mechanical fault diagnosis \cite{Stefani2009, Liu2020, Zhao2020a, Tang2021, Li2022e}, methods based on handcrafted features are widely applied. 
These methods typically first employ \emph{time-frequency analysis} (TFA) algorithms or other data preprocess approaches \cite{Huang2019} aiming to mitigate the influence caused by condition variations.
Common TFA methods include \emph{Fourier transform} (FT) \cite{Wang2024b, Liu2024a}, \emph{wavelet transform} (WT) \cite{Gritli2011}, \emph{synchro extracting transform} (SET) \cite{Ma2022, Wu2023, Yu2023}, etc.
Subsequently, the extracted features are used to identify specific faults using either manual approaches or machine learning-based classifiers \cite{Shan2023}. 
The structural schematic of the handcrafted feature-based framework is illustrated in Fig. \ref{handcrafted}.

Qin \emph{et al.} proposed an \emph{adaptive fast chirplet transform} (AFCT) based on the adaptive optimal search angle band to effectively extract fault features under time-varying speed condition \cite{Qin2023}.
Wang \emph{et al.} proposed a sparse and low-rank decomposition of the time-frequency representation technique to effectively capture the underlying characteristic frequencies of bearings under variable speed conditions\cite{Wang2022b}.
While in \cite{Ding2024}, Ding \emph{et al.} put forward a low-rank constrained \emph{multi-kernel general parameterized TF transform} (MKGPTFT) to generate high-quality time-frequency plane for accurate fault diagnosis.
In \cite{Ying2024}, an angle-time double-layer decomposition structure termed \emph{order-frequency Holo-Hilbert spectral analysis} (OFHHSA) is established to reveal the interaction relationship between time-based carriers and angle-based modulations.
An \emph{adaptive time frequency extraction mode decomposition} (ATFEMD) method was proposed in \cite{Huo2024} to solve the problems of time-frequency energy lack of concentration, poor robustness of instantaneous frequency extraction, and mode aliasing in signal decomposition. It uses ridge extraction to capture the distinctive time-frequency information within the time-frequency distribution, ensuring that the model can be well adapted to realistic scenarios.
By combining EMD and \emph{adaptive time-varying parameter short-time Fourier synchronous squeezing transform} (AFSST), Wei \emph{et al.} proposed a bearing fault diagnosis method \cite{Wei2022}. Compared with traditional methods, this approach can obtain higher frequency resolution at a lower sampling rate.
In \cite{Song2023}, the multispectral lossless preprocessing module was established to eliminate the influence of variable rotating speeds and avoid the loss of fault information.
Schmidt \emph{et al.} proposed a systematic framework for obtaining consistent feature planes under time-varying operating conditions for gearbox fault diagnosis \cite{Schmidt2021}.

In addition, \emph{order tracking} (OT) is also a common data processing method in MCFD.
In \cite{Sapena-Bano2017}, an extension to non-stationary conditions of the \emph{harmonic order tracking approach} (HOTA) was introduced. This method can obtain condition-invariant patterns, allowing nonskilled personnel to perform reliable diagnostics and simplifying the development of automated diagnostic methods based on machine learning algorithms.
While in \cite{Wang2017}, a hybrid approach is proposed for diagnosing faults in roller bearings under variable speed conditions. This method integrates \emph{computed order tracking} (COT) with a \emph{variational mode decomposition} (VMD)-based time-frequency representation to capture the relationship between the angular information of the vibration signal and the fault characteristic order.
Wang \emph{et al.} proposed an IEWT-based enhanced envelope order spectrum method, which integrates the advantages of the \emph{improved empirical wavelet transform} (IEWT), COT, and \emph{singular value ratio spectrum} (SVRS) denoising technology to achieve reliable and efficient fault diagnosis under variable speed conditions \cite{Wang2019}.
In \cite{Hu2024}, Hu \emph{et al.} proposed a \emph{tacholess order tracking} method (TLOT) based on ridge extraction method to address the challenge of non-stationary vibration signals due to varying rotational speed conditions.

\subsection{Learned representation-based framework}
\begin{figure}[!htpb]
	\centering
	\includegraphics[width=2.2in, keepaspectratio]{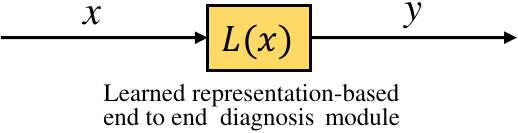}
	\caption{Structural schematic of the learned representation-based framework.}
	\label{learned}
\end{figure}

\begin{figure}[htpb]
	\centering
	\includegraphics[width=3.3in, keepaspectratio]{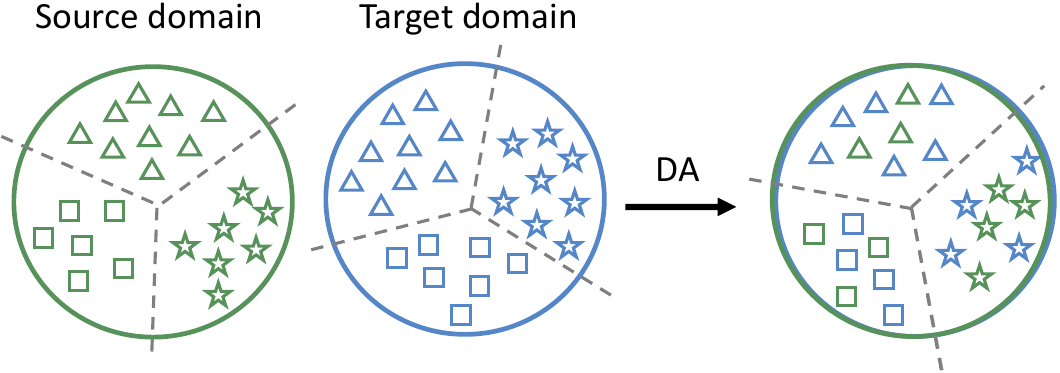}
	\caption{Schematic diagrams of domain adaption.}
	\label{DA}
\end{figure}

Learned representation-based methods typically employ deep learning approaches to extract fault-invariant features, aiming to develop a generalized model applicable across different operating conditions. The schematic diagram of its framework is shown in Fig. \ref{learned}.
One common strategy within this category is \emph{domain adaption} (DA) \cite{Xia2022, Han2023, Lei2021a,Yang2022, Qian2019, Choudhary2023,Wu2022, Tong2018, Wu2022a, Wang2022a}. 
The core idea is to initially train a model using data from the source domain, followed by fine-tuning the model with a subset of data from the target domain \cite{Yang2023}.
By aligning the data distributions across different domains, knowledge from the source domain can be transferred to the target domain, thereby ensuring accurate fault diagnosis in the target domain \cite{An2019a, Yu2022, An2019, Qian2021, Yao2023, Zhang2018, Lei2021}, as illustrated in Fig. \ref{DA}.


In \cite{Song2024}, a \emph{convolutional neural networks-bidirectional long short-term memory} (CNN-BiLSTM) network combined with improved \emph{particle swarm optimization} (PSO) was developed for precise bearing fault diagnosis. The method leverages temporal features and transfers pre-trained models to new conditions, addressing data scarcity in varied working environments.
\cite{Wang2021} introduced a deformable CNN-DLSTM model integrated with transfer learning strategies to improve bearing fault diagnosis under multiple operating conditions. This method enables the model to learn transferable features, achieving robust performance with minimal data from new conditions.
Shao \emph{et al.} proposed a modified transfer CNN driven by thermal images, which incorporates stochastic pooling and LRelU to achieve high adaptability \cite{Shao2021}.
Su \emph{et al.} introduced a \emph{dilated convolution deep belief network-dynamic multi-layer perceptron} (DCDBN-DMLP) which utilizes a multi-layer \emph{maximum mean discrepancy} (MMD) technique to reduce distribution discrepancy \cite{Su2022a}.
An \emph{et al.} proposed an \emph{unsupervised contrast domain adaptive network} (UCDAN) for cross-domain bearing fault diagnosis \cite{An2022}. It uses contrast estimation terms to increase the distance between samples from different classes.
In \cite{Kuang2022}, an end-to-end prototype-guided bi-level adversarial domain adaptation network is proposed to narrow the domain shift due to different operating conditions.
A \emph{prototype and stochastic neural network-based twice adversarial domain adaption} (PSNN-TADA) was proposed to effectively align category-wise fault features \cite{Yang2024}. This method adopts a stochastic neural network-based classifier to mitigate the misaligned fault features in the target domain, ensuring more discriminative decision boundaries.
\cite{Zhai2022} and \cite{Hu2020} used \emph{adaptive batch normalization} (AdaBN) related strategies to enhance the model adaptation to the distribution of the target domain.

\begin{figure}[htpb]
	\centering
	\includegraphics[width=3.3in, keepaspectratio]{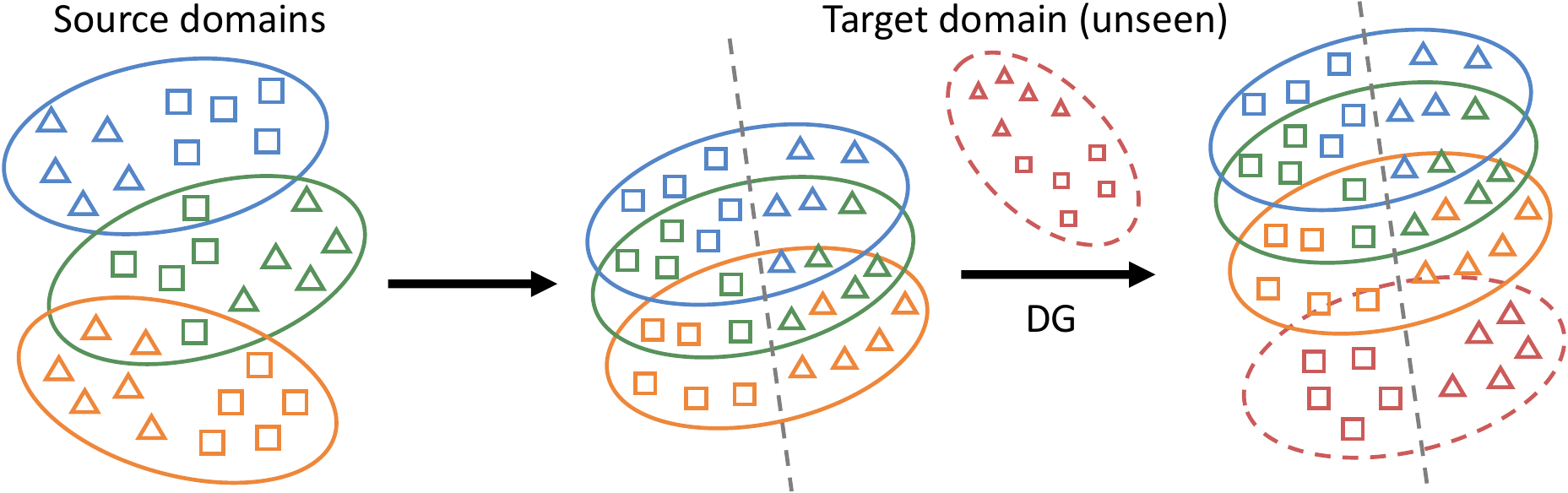}
	\caption{Schematic diagrams of domain generalization.}
	\label{DG}
\end{figure}

Although the aforementioned methods have demonstrated promising results in specific application scenarios, they are highly dependent on precise condition identification \cite{Xu2020, Shi2025}.
In practical industrial scenarios, unknown conditions frequently arise due to unpredictable variations in operating conditions \cite{Tang2022a, Lei2023, Xu2022}.
In this context, \emph{domain generalization} (DG) has emerged as a promising approach to tackle the challenge of unknown operating conditions in MCFD \cite{Ren2023, Tang2024}.
In the absence of information from the target domain, these methods construct a model using the available condition data to mitigate the impact of varying conditions.
When new samples arrive, no further adaptation is required, and satisfactory results can be obtained by directly applying the offline model.
The core idea is to extract diverse domain-invariant representations and enhance the differentiation within the feature space. This framework makes the model more robust to out-of-distribution data, allowing it to generalize effectively to previously unseen operating conditions \cite{Shi2023a, Wang2023, Liu2024}.
The schematic diagrams of domain generalization are presented in Fig. \ref{DG}.

In \cite{Wang2024}, a \emph{ProbSparse attention-based transformer} (PSAT) for in situ fault diagnosis is proposed, which reduces the model complexity and storage cost of diagnostic samples.
Han \emph{et al.} proposed a hybrid generalization network termed the \emph{intrinsic and extrinsic domain generalization network} (IEDGNet) to solve the diagnosis problem under unseen operating conditions\cite {Han2021a}. It uses the triplet loss minimization of the intrinsic multi-source data to reduce the distribution change caused by varying operating conditions, ensuring robustness and generalization ability.
In \cite{Li2023b}, Li \emph{et al.} introduced a \emph{cross-domain augmentation} (CDA) approach for fault diagnosis under unseen operating conditions. By using an \emph{adversarial domain-augmented generalization} (ADAG) method with convex combinations of feature-label pairs, this approach promotes model generalization by learning domain-invariant features across multi-source and augmented domains.
To tackle limited sample availability in fault diagnosis, Zheng \emph{et al.} introduced a meta-learning approach with adaptive input and attention mechanisms \cite{Zheng2023}, enhancing feature extraction and generalization capabilities.
Similarly, Li \emph{et al.} also proposed a meta-learning method for bearing fault diagnosis under complex operating conditions with limited data \cite{Li2021a}. By converting raw signals into time-frequency images and applying meta-learning protocols, it leverages prior knowledge from multiple tasks to achieve fast and accurate few-shot learning for unseen conditions.
To address the issue of limited labeled samples, a mutual-assistance network for \emph{semi-supervised domain generalization fault diagnosis} (SemiDGFD) is proposed in \cite{Zhao2023a}. This method employs a pseudo-label technique and uses an entropy-based purification mechanism to enhance pseudo-label quality.
In \cite{Shi2023b}, Shi \emph{et al.} proposed a \emph{reliable feature-assisted contrastive generalization net} (RFACGN) to address the challenges of lack of explanation in intelligent fault diagnosis. The method uses a contrastive framework to minimize domain-specific knowledge and a multi-branch module to focus on fault-related features while also introducing a confidence metric to evaluate the reliability of the results.
Wang \emph{et al.} proposed an \emph{enhanced transformer with asymmetric loss function} (ETALF) approach to tackle few-shot fault diagnosis with noisy labels \cite{Wang2024a}. This method enhances robustness to label noise by dynamically measuring fault sample similarities and using an asymmetric loss function to improve diagnostic accuracy.
Liu \emph{et al.} proposed an \emph{information-induced feature decomposition and augmentation} framework (IIFDA) to address bearing fault diagnosis under non-stationary conditions \cite{Liu2023}.
The work discussed in \cite{Zhao2020, Wang2021a} tackles the data imbalance problem under various operating conditions. These two methods utilize \emph{batch normalization} (BN) to mitigate the distribution difference between the training dataset and the testing dataset.

\begin{table*}[!htbp]
	\begin{center}	
		\centering
		\caption{Summary of Recent Typical Single-model-based MCFD Approaches}\label{Table_1}
		\setlength{\tabcolsep}{5.5mm}{
			\begin{tabular}{{c}{l}{c}{c}{c}{c}{c}{c}{c}{c}{c}{c}{c}{c}{c}{c}{c}{c}p{1cm}}
				\specialrule{0.1em}{1pt}{1pt}
				\specialrule{0.1em}{1pt}{3pt}
				\cmidrulewidth = 0.5em
				\multirow{2.5}{*}{Year} & \multirow{2.5}{*}{Literature} & \multirow{2.5}{*}{Technical roadmap} & \multicolumn{2}{c}{Main issues} & \multirow{2.5}{*}{System object}\\ \cmidrule(r){4-5}
				
				& & & steady & unsteady & &  \\
				\specialrule{0.1em}{1pt}{3pt}
				
				2013 & Xue \emph{et al.} \cite{Xue2013} & handcrafted feature & &\checkmark & motor \\
				2017 & Sapena \emph{et al.} \cite{Sapena-Bano2017} & handcrafted feature & &\checkmark & motor \\
				2017 & Zhang \emph{et al.} \cite{Zhang2017} & learned representation & \checkmark & & bearing \\
				2018 & Chen \emph{et al.} \cite{Chen2018} & handcrafted feature & \checkmark & & gearbox \\
				2019 & Hasan \emph{et al.} \cite{Hasan2019} & learned representation & \checkmark & & bearing \\
				2019 & Li \emph{et al.} \cite{Li2019a} & handcrafted feature & &\checkmark & gearbox \\
				2020 & Singh \emph{et al.} \cite{Singh2020} & learned representation & \checkmark & & gearbox \\
				2020 & Wu \emph{et al.} \cite{Wu2020} & learned representation & \checkmark & & chemical processes \\
				2020 & Peng \emph{et al.} \cite{Peng2020} & learned representation & \checkmark & & bearing \\
				2020 & Gu \emph{et al.} \cite{Gu2020} & handcrafted feature & &\checkmark & bearing \\
				2021 & Jahagirdar \emph{et al.} \cite{Jahagirdar2021} & handcrafted feature & &\checkmark & bearing \\
				2021 & Schmidt \emph{et al.} \cite{Schmidt2021} & handcrafted feature & &\checkmark & gearbox \\
				2021 & Han \emph{et al.} \cite{Han2021} & learned representation & \checkmark & & bearing \\
				2021 & Mao \emph{et al.} \cite{Mao2021} & learned representation & \checkmark & & bearing \\
				2021 & Li \emph{et al.} \cite{Li2021b} & learned representation & \checkmark & & bearing \\
				2021 & Chen \emph{et al.} \cite{Chen2021a} & handcrafted feature & &\checkmark & gearbox \\
				2022 & Atta \emph{et al.} \cite{Atta2022} & handcrafted feature & &\checkmark & bearing \\
				2022 & Karabacak \emph{et al.} \cite{Karabacak2022} & handcrafted feature & \checkmark & & gearbox \\
				2022 & Kavianpour \emph{et al.} \cite{Kavianpour2022} & learned representation & \checkmark & & bearing \\
				2022 & Park \emph{et al.} \cite{Park2022} & handcrafted feature & &\checkmark & motor \\
				2022 & Su \emph{et al.} \cite{Su2022} & learned representation & \checkmark & & bearing \\
				2022 & Xu \emph{et al.} \cite{Xu2022a} & learned representation & &\checkmark & motor \\
				2022 & Li \emph{et al.} \cite{Li2022} & handcrafted feature & &\checkmark & bearing \\
				2022 & Li \emph{et al.} \cite{Li2022d} & learned representation & \checkmark & & nuclear power plant \\
				2023 & Shi \emph{et al.} \cite{Shi2023} & learned representation & \checkmark & & bearing \\
				2023 & An \emph{et al.} \cite{An2023} & learned representation & \checkmark & & bearing \\
				2023 & Chen \emph{et al.} \cite{Chen2023b} & learned representation & \checkmark & & bearing \\
				2023 & Ren \emph{et al.} \cite{Ren2023} & learned representation & \checkmark & & bearing \\
				2023 & Chen \emph{et al.} \cite{Chen2023c} & learned representation & \checkmark & & bearing \\
				2023 & Liu \emph{et al.} \cite{Liu2023} & learned representation & &\checkmark & bearing \\
				2023 & Jiang \emph{et al.} \cite{Jiang2023} & learned representation & \checkmark & & chillers \\
				2023 & Zhang \emph{et al.} \cite{Zhang2023a} & handcrafted feature & &\checkmark & gearbox \\
				2023 & Bai \emph{et al.} \cite{Bai2023} & handcrafted feature & \checkmark & & bearing \\
				2023 & Yu \emph{et al.} \cite{Yu2023a} & learned representation & \checkmark & & bearing \\
				2023 & Ding \emph{et al.} \cite{Ding2023} & learned representation & \checkmark & & bearing \\
				2023 & Li \emph{et al.} \cite{Li2023} & learned representation & \checkmark & & bearing and gear \\
				2023 & Zhang \emph{et al.} \cite{Zhang2023} & learned representation & \checkmark & & bearing \\
				2023 & Zhu \emph{et al.} \cite{Zhu2023} & learned representation & \checkmark & & bearing and rotor \\
				2024 & Shi \emph{et al.} \cite{Shi2024} & learned representation & \checkmark & & bearing \\
				2024 & Wang \emph{et al.} \cite{Wang2024} & learned representation & \checkmark & & industrial robots \\
				2024 & Fang \emph{et al.} \cite{Fang2024} & handcrafted feature & \checkmark & & compressors \\
				2024 & Mao \emph{et al.} \cite{Mao2024} & learned representation & \checkmark & & satellite \\
				2024 & Che \emph{et al.} \cite{Che2024} & learned representation & \checkmark & & bearing \\
				2024 & Zhang \emph{et al.} \cite{Zhang2024} & learned representation & \checkmark & & gearbox \\

				\specialrule{0.1em}{3pt}{1pt}
				\specialrule{0.1em}{1pt}{1pt}
		\end{tabular}}\label{single_table}
	\end{center}
	
\end{table*}

\section{multi-model-based MCFD methods}
The multi-model approach can be further divided into two categories: fusion-guided and identification-guided methods.
The advantage of fusion-guided methods lies in their ability to integrate the results from multiple sub-models, thereby fully mining and utilizing the available information. 
However, the diagnostic accuracy of this approach may be negatively affected by erroneous sub-models. 
In contrast, identification-guided methods only invoke the sub-model corresponding to the current operating condition. 
If the condition identification is accurate, this method can theoretically achieve optimal performance. 
However, because the sub-models are trained on offline data, this approach is less adaptive when encountering unknown operating conditions.
In this section, we will provide a detailed review of the existing literature based on these two frameworks.

\begin{table*}[!htbp]
	\begin{center}
		\centering
		\caption{Summary of Recent Typical Multi-model-based MCFD Approaches}\label{Table_2}
		\setlength{\tabcolsep}{5mm}{
			\begin{tabular}{{c}{l}{c}{c}{c}{c}{c}{c}{c}{c}{c}{c}{c}{c}{c}{c}{c}{c}p{1cm}}
				\specialrule{0.1em}{1pt}{1pt}
				\specialrule{0.1em}{1pt}{3pt}
				\cmidrulewidth = 0.5em
				\multirow{2.5}{*}{Year} & \multirow{2.5}{*}{Literature} & \multirow{2.5}{*}{Technical roadmap} & \multicolumn{2}{c}{Main issues} & \multirow{2.5}{*}{System object}\\ \cmidrule(r){4-5}
				
				& & & steady & unsteady & &  \\
				\specialrule{0.1em}{1pt}{3pt}
				
				2013 & Yu \emph{et al.} \cite{Yu2013} & fusion-guided & \checkmark & & chemical process \\
				2016 & Jiang \emph{et al.} \cite{Jiang2016} & identification-guided & \checkmark & & chemical process \\
				2018 & Chen \emph{et al.} \cite{Chen2018a} & identification-guided & \checkmark & & electrical traction systems \\
				2018 & Ma \emph{et al.} \cite{Ma2018} & fusion-guided & \checkmark & & modern hot strip mill process \\
				2019 & Shang \emph{et al.} \cite{Shang2019} & identification-guided & \checkmark & & sensor \\
				2020 & Song \emph{et al.} \cite{Song2020} & fusion-guided & \checkmark & & chemical process \\
				2020 & He \emph{et al.} \cite{He2020} & fusion-guided & \checkmark & & bearing and gear \\
				2022 & Liang \emph{et al.} \cite{Liang2022} & identification-guided & \checkmark & & fixed-wing unmanned aerial vehicles \\
				2022 & Ye \emph{et al.} \cite{Ye2022} & fusion-guided & \checkmark & & rotor and bearing \\
				2023 & Xu \emph{et al.} \cite{Xu2023} & fusion-guided & \checkmark & & gearbox \\
				2023 & Li \emph{et al.} \cite{Li2023a} & fusion-guided & \checkmark & & chemical process \\
				2023 & Zhou \emph{et al.} \cite{Zhou2023} & fusion-guided & \checkmark & & chemical process \\
				2023 & Liu \emph{et al.} \cite{Liu2023a} & fusion-guided & \checkmark & & chemical process \\
				2024 & Chen \emph{et al.} \cite{Chen2024} & fusion-guided & \checkmark & & bearing \\
				2024 & Gao \emph{et al.} \cite{Gao2024} & fusion-guided & \checkmark & & bearing \\

				\specialrule{0.1em}{3pt}{1pt}
				\specialrule{0.1em}{1pt}{1pt}
		\end{tabular}}	\label{multi_table}
	\end{center}
\end{table*}

\subsection{Fusion-guided framework}
\begin{figure}[!htpb]
	\centering
	\includegraphics[width=2.2in, keepaspectratio]{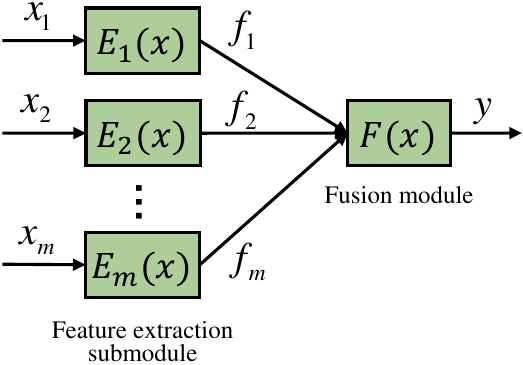}
	\caption{Structural schematic of the fusion-guided framework.}
	\label{fusion}
\end{figure}

Given the availability of data from multiple operating conditions during the offline stage, directly constructing a single diagnostic model may lead to performance degradation due to the interference between condition information and fault information. To tackle this issue, the fusion-guided framework develops individual sub-models for each operating condition using offline-collected data.
In the online stage, incoming samples are processed by each sub-model, and the outputs are combined using a decision fusion strategy to generate the final fault diagnosis result. 
This type of approach makes efficient use of data from various operating conditions, aiming to reduce information loss \cite{Zhou2023, Chen2024}. The structural schematic of the fusion-guided framework is illustrated in Fig. \ref{fusion}.

In \cite{Ye2022}, Ye \emph{et al.} proposed a \emph{deep negative correlation multisource domain adaptation network} (DNC-MDAN). This method integrates features adapted from multiple source domains into a DNC-based ensemble classifier, where the final result is obtained by averaging the outputs of the sub-classifiers.
\cite{He2020} proposed an ensemble transfer convolutional neural networks approach. In this method, a corresponding CNN is trained on each source domain and fine-tuned with target domain data. A decision fusion strategy is then designed to integrate the predictions from all CNNs, producing a comprehensive result.
\cite{Xu2023} proposed a \emph{decision self-regulating network} (DSRN) that constructs multiple CNN sub-classifiers and employs a score unit to regulate the diagnostic results of each sub-model under specific operating conditions. The proposed method also addresses the issue of class imbalance in fault diagnosis tasks using multi-classifiers and ensemble learning algorithms.
An \emph{evidential ensemble preference-guided learning} (EEPL) approach is proposed in \cite{Liu2023a}, which utilizes recursive \emph{evidential reasoning} (ER) technology to integrate information from base classifiers within an ensemble-based \emph{broad learning system} (BLS). Additionally, this method incorporates an iterative parameter updating mechanism, enabling real-time adjustments according to varying operating conditions.
\cite{Yu2013} proposed a \emph{Bayesian inference-based Gaussian mixture contribution} (BIGMC) index, which integrates multiple local contribution indices into the global contribution index. This index can be used to identify key fault variables. Building on this, \cite{Ma2018} introduced the \emph{Bayesian inference-based robust Gaussian mixture contribution} (BIRGMC) index, which takes into account the relationship between various industrial operating conditions and comprehensive quality-related faults.
In \cite{Li2023a}, Li \emph{et al.} proposed a \emph{feature-level and class-level multisource domain adaptation approach} (FC-MSDA), which integrates multiple predictions from domain-specific classifiers via an information fusion module. This fusion module leverages the similarity of shared features between the source and target domains.

\subsection{Identification-guided framework}
\begin{figure}[!htpb]
	\centering
	\includegraphics[width=2.2in, keepaspectratio]{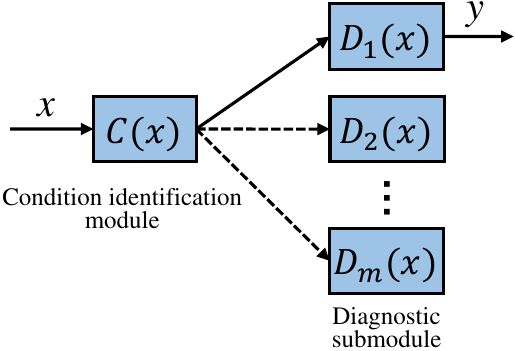}
	\caption{Structural schematic of the identification-guided framework.}
	\label{identification}
\end{figure}

Similarly, the identification-guided MCFD method also builds submodels for each known operating condition during the offline stage.
The main difference from the fusion-guided framework is that, during the online phase, this method first identifies the operating condition and then selects the corresponding submodel to perform fault diagnosis, as shown in Fig. \ref{identification}. 

\cite{Jiang2016} employed a condition identification method based on Bayesian inference, utilizing independent local \emph{principal component analysis} (PCA) models for each condition to eliminate collinearity among variables and extract latent factors while also applying a \emph{genetic algorithm} (GA) for \emph{optimal principal component} (OPC) selection. Then, a Bayesian diagnosis system based on OPC is constructed to diagnose faults.
\cite{Chen2018a} established individual PCA models for each operating condition during the offline stage. Once the operating condition is identified during the online stage, the corresponding sub-model is applied to reduce the dimensionality of the samples. This facilitates fault detection and isolation, thereby enabling accurate diagnosis of incipient multi-condition sensor faults in the electrical traction system.
\cite{Shang2019} proposed a \emph{condition-independent Bayesian learning-based recursive transformation component statistical analysis} (CIBL-RTCSA) method for addressing incipient sensor fault in multi-operating conditions. 
During the offline stage, this method independently calculates reference statistics for each condition. In the online phase, a window-switching-based multiple RTCSA approach is employed after performing condition identification to achieve enhanced fault detection and isolation performance.
In \cite{Liang2022}, Liang \emph{et al.} proposed a data-driven framework for fault diagnosis in fixed-wing unmanned aerial vehicles under multiple operating conditions. The framework employs DBSCAN and KNN algorithms based on \emph{shared nearest neighbor based distance} (SNND) for offline condition classification and online condition recognition. Once the operating condition is identified, a \emph{multiple condition oriented dynamic KPCA} (MDKPCA) algorithm is applied for fault diagnosis.

\section{Evaluation Metric}
Evaluation metrics are of great significance in evaluating the diagnostic performance of MCFD systems by offering quantifiable standards for measurement. 
Since the MCFD task aims to identify specific fault types, it is typically regarded as a multi-class classification problem.
In the subsequent descriptions, for an MCFD task with $m$ types of faults, \emph{true positive} (TP) refers to the number of samples correctly predicted as positive; \emph{true negative} (TN) refers to the number of samples correctly predicted as negative; \emph{false positive} (FP) refers to the number of samples incorrectly predicted as positive; and \emph{false negative} (FN) refers to the number of samples incorrectly predicted as negative.
The rest of this section outlines several commonly used evaluation metrics for tackling the MCFD problem.

\subsection{False Alarm Rate}
\emph{False Alarm Rate} is a critical performance metric in fault diagnosis, reflecting the percentage of samples that are mistakenly classified as positive cases.
A high false alarm rate can lead to unnecessary operational and maintenance costs and may result in frequent equipment shutdowns in non-fault scenarios, affecting production efficiency. 
Therefore, minimizing the false alarm rate is crucial for enhancing the reliability and practicality of a fault diagnosis system.
The definition is as follows:
\begin{equation}
	\text{FAR}=\frac{\text{FP}}{\text{FP}+\text{TN}}.
\end{equation}

\subsection{Missing Alarm Rate}
\emph{Missing Alarm Rate} represents the proportion of actual fault instances the system fails to identify.
Specifically, it refers to the proportion of samples incorrectly classified as negative. 
A high missing alarm rate indicates that critical fault information might be overlooked, significantly increasing the risk of accidents or system failures, which can potentially lead to catastrophic consequences.
The definition is as follows:
\begin{equation}
	\text{MAR}=\frac{\text{FN}}{\text{TP}+\text{FN}}.
\end{equation}

\subsection{Accuracy}
\emph{Accuracy} measures the proportion of correctly predicted samples out of the total samples tested. 
It is one of the most commonly used and convenient evaluation metrics for assessing diagnostic performance. Its definition is as follows:
\begin{equation}
	\text{Accuracy}=\frac{\text{TP}+\text{TN}}{\text{TP}+\text{FN}+\text{TN}+\text{FP}}.
\end{equation}

\subsection{G-mean}
\emph{G-mean} computes the geometric mean of sensitivity (Recall) and specificity. It effectively reduces the bias introduced by class imbalance in the evaluation process by calculating the Recall for each class individually and applying appropriate weighting.
Therefore, it serves as a widely adopted evaluation metric for addressing imbalanced diagnostic tasks. 
The specific definition is as follows:
\begin{equation}
	\text{G-mean}=\left(\prod_{i=1}^{m}\frac{\text{TP}_i}{\text{TP}_i+\text{FN}_i}\right)^{\frac{1}{m}}.
\end{equation}

\subsection{Macro-F1-score}
\emph{Macro-F1-score} is used to evaluate the overall performance of a model in balancing accuracy and coverage, especially in scenarios with imbalanced sample classes. 
For the $i$-th class, the Precision and Recall values can be calculated as follows:
\begin{equation}
	\text{Precision}_i=\frac{\text{TP}_i}{\text{TP}_i+\text{FP}_i},
\end{equation}
and
\begin{equation}
	\text{Recall}_i=\frac{\text{TP}_i}{\text{TP}_i+\text{FN}_i}.
\end{equation}
Building on this, the Macro Average Precision and Recall are determined by taking the arithmetic mean of the metrics for each individual class.
\begin{equation}
	\text{Macro-Precision}=\frac{\sum_{i=1}^{m}\text{Precision}_i}{m},
\end{equation}
and
\begin{equation}
	\text{Macro-Recall}=\frac{\sum_{i=1}^{m}\text{Recall}_i}{m}.
\end{equation}
The Macro-F1-Score is then obtained by calculating the harmonic mean of Macro-Precision and Macro-Recall, as illustrated below:
\begin{equation}
	\text{Macro-F1-score}=\frac{2\times\text{Macro-Precision}\times\text{Macro-Recall}}{\text{Macro-Precision}+\text{Macro-Recall}}. \label{macro-F1}
\end{equation}

\section{Typical real-world application}
MCFD has been widely applied in various domains, such as mechanical systems, chemical systems, energy systems, and satellite systems, as shown in Tables \ref{single_table} and \ref{multi_table}. 
Due to the distinct data characteristics of different diagnostic targets, the challenges posed to the diagnostic models also vary.
In this section, we will provide a brief overview of the two main application scenarios of MCFD: mechanical systems and chemical systems.

\subsection{Mechanical System}
\begin{figure}[htpb]
	\centering
	\includegraphics[width=3.2in, keepaspectratio]{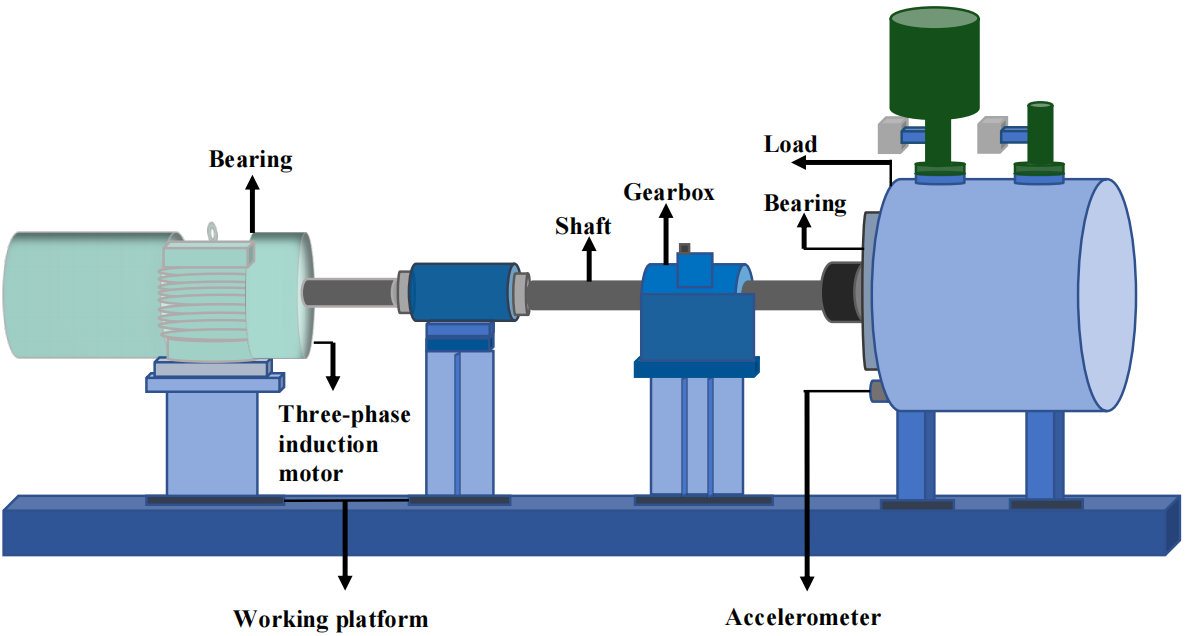}
	\caption{An example of a typical mechanical system experimental platform \cite{10292791}.}
	\label{Mechanical}
\end{figure}

Mechanical components are essential elements of industrial equipment. Their operational status directly affects the reliability and safety of the entire system \cite{Lu2023b, Sohaib2020}. A typical mechanical system is composed of multiple critical components, such as bearings, rotors, and gearboxes, as shown in Fig. \ref{Mechanical}. The cooperative functioning of these components determines the overall performance of the equipment. It also significantly influences its service life and maintenance costs. 

Faults in mechanical systems not only disrupt production processes but also pose significant safety risks to personnel, potentially leading to equipment damage, financial losses, and even casualties \cite{Ge2024, Shen2020a, An2023a}.
Therefore, it is crucial to implement accurate and efficient fault diagnosis for mechanical systems. 
To achieve this, vibration sensors are installed on critical components to collect vibration signals generated during production. 
With advancements in sensor technology, the sampling frequency and sensitivity of sensors have significantly improved. 
This progress enables the acquisition of large volumes of high-quality data in real-time \cite{Chen2023a}. 
These data are typically annotated by experts to construct offline datasets, which provide a solid foundation for training fault diagnosis models.
For rotating machinery and other mechanical systems, changes in speed and load are common. These variations are caused by production fluctuations or environmental disturbances. Such variations create different operating conditions, leading to significant shifts in data distribution. These shifts reduce the performance of existing fault diagnosis models and pose challenges for maintaining diagnostic accuracy under varying conditions.

Tables \ref{Table_1} and \ref{Table_2} summarize the relevant literature in the field of MCFD in recent years. 
It is observed that most existing methods rely on single-model approaches. 
The handcrafted feature-based method is one of the commonly adopted frameworks \cite{Miao2021, Tian2015, Fu2023}. These methods utilize techniques such as time-frequency analysis or other data processing strategies to remove condition-related features \cite{Chen2023, Li2019, Wang2018}.
By doing so, fault-invariant features are extracted from vibration signals, which helps to improve diagnostic accuracy.
In recent years, the rapid development of deep learning technologies has provided new solutions for MCFD \cite{Li2024a, Li2022c, Sun2023}. Transfer learning has become a widely used approach for addressing the challenges associated with condition variations \cite{Shen2021, Wu2020a, Zhu2020, Chen2020, Zhang2019}. Techniques such as domain adaptation and domain generalization have been employed to reduce the dependency on large-scale labeled datasets. At the same time, these techniques enhance the generalization ability of models across different operating conditions. This progress offers new possibilities for fault diagnosis in mechanical systems under complex industrial environments and contributes to advancements in this research field.
We have also summarized the representative multi-condition mechanical datasets used in studies, as illustrated in Table \ref{dataset_table}.
\begin{table*}[htbp]
	\centering
	\caption{Summary of Typical Multiple Operating Conditions Mechanical Datasets}
	\setlength{\tabcolsep}{1.1mm}{
		\begin{tabular}{lcccccccccc}
			\toprule
			\toprule
			\multirow{2}[3]{*}{Name} & \multicolumn{3}{c}{Fault} & \multicolumn{2}{c}{Operating condition} & \multicolumn{3}{c}{Sampling} & \multicolumn{1}{c}{\multirow{1}[4]{*}{Operational}} & \multirow{2}[4]{*}{Channels} \\
			\cmidrule(r){2-4}\cmidrule(r){5-6}\cmidrule(r){7-9}          & types & levels & compound fault & non-stationary & stationary & variables & frequency & period &  variables     &  \\
			\midrule
			MCC5-THU \cite{MCC5THU} & 7     & 3     & 2     & 48    & 24    & vibration & 12.8kHz & 60s   & speed, load & 6 \\
			\midrule
			\multirow{2}[2]{*}{CWRU \cite{CWRU}} & 3     & 3     & \textbackslash{} & \textbackslash{} & 4     & vibration & 12kHz & \textbackslash{} & speed, load & 3 \\
			\cmidrule{2-11}          & 3     & 4     & \textbackslash{} & \textbackslash{} & 4     & vibration & 48kHz & \textbackslash{} & speed, load & 3 \\
			\midrule
			Ottawa \cite{Ottawa} & 2     & \textbackslash{} & \textbackslash{} & 6     & \textbackslash{} & vibration & 200kHz & 10s   & speed & 2 \\
			\midrule
			\multirow{5}[4]{*}{KAIST \cite{KAIST}} & \multirow{3}[2]{*}{4} & \multirow{3}[2]{*}{3} & \multirow{3}[2]{*}{\textbackslash{}} & \multirow{3}[2]{*}{\textbackslash{}} & \multirow{3}[2]{*}{3} & vibration & \multirow{3}[2]{*}{25.6kHz} & \multirow{3}[2]{*}{60s} & \multirow{3}[2]{*}{load} & 4 \\
			&       &       &       &       &       & current &       &       &       & 3 \\
			&       &       &       &       &       & temperature &       &       &       & 1 \\
			\cmidrule{2-11}          & \multirow{2}[2]{*}{3} & \multirow{2}[2]{*}{\textbackslash{}} & \multirow{2}[2]{*}{\textbackslash{}} & \multirow{2}[2]{*}{1} & \multirow{2}[2]{*}{1} & vibration & 25.6kHz & \multirow{2}[2]{*}{2700s} & \multirow{2}[2]{*}{speed} & 4 \\
			&       &       &       &       &       & current & 100kHz &       &       & 3 \\
			\midrule
			PU \cite{PUB} & 2     & 5     & 3     & \textbackslash{} & 4     & current, vibration & 64kHz & 4s    & speed, load, radial force & 20 \\
			\midrule
			PBTU \cite{PBTU} & 2     & 3     & \textbackslash{} & \textbackslash{} & 21    & vibration & 51.2kHz & 10s   & speed, load & 6 \\
			\midrule
			XJTU-SY \cite{XJTUSY} & 3     & \textbackslash{} & 2     & \textbackslash{} & 3     & vibration & 25.6kHz & \textbackslash{} & speed, radial force & 2 \\
			\midrule
			JNU \cite{JNU} & 3     & \textbackslash{} & 1     & \textbackslash{} & 3     & vibration & 50kHz & 20s   & speed, load & 1 \\
			\bottomrule
			\bottomrule
	\end{tabular}}
	\begin{tablenotes}
		\item[]Notes: The "Fault types" section only counts the number of faults in the dataset, excluding healthy conditions;\\
		\item[]The \textbackslash{} symbol indicates that the information is either not mentioned in the referenced literature or is inconvenient to present.
	\end{tablenotes}
	\label{dataset_table}%
\end{table*}%

\subsection{Chemical System}
\begin{figure}[htpb]
	\centering
	\includegraphics[width=3.4in, keepaspectratio]{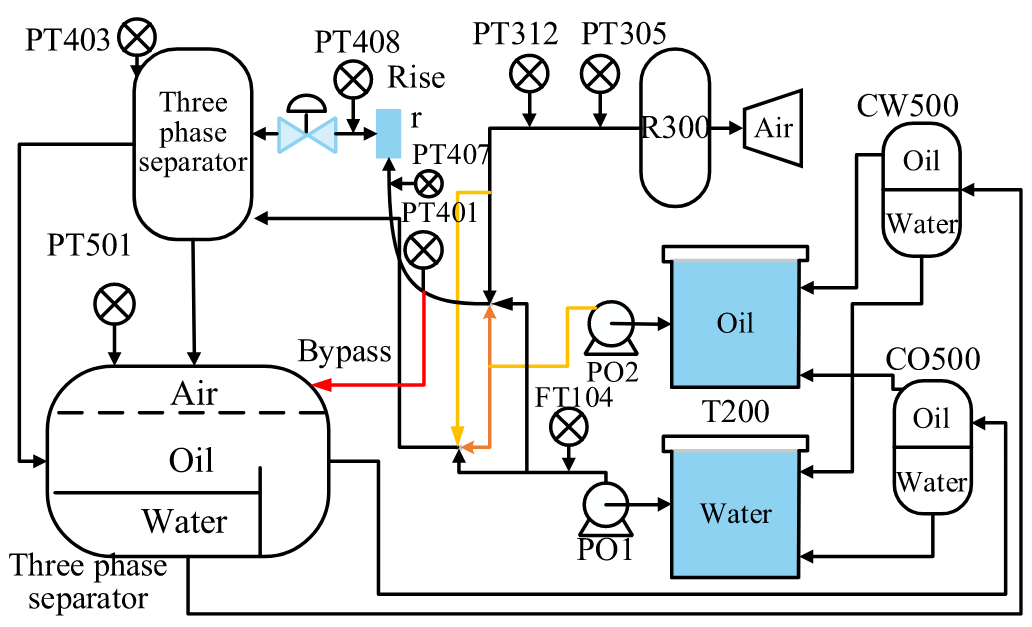}
	\caption{An example of a typical chemical system experimental platform \cite{Li2023a}.}
	\label{Chemical}
\end{figure}

The issue of multiple operating conditions in chemical systems has become a prominent research focus in recent years. 
Chemical systems often involve intense chemical reactions, which can be accompanied by substantial heat release or the generation of toxic by-products. A typical experimental platform for chemical systems is shown in Fig. \ref{Chemical}.
In the event of fault occurrences, such reactions can lead to catastrophic consequences, such as hazardous substance leaks or explosions. 
Consequently, compared to other industrial processes, faults in chemical systems are more likely to result in severe safety incidents. 
The inherent complexity of chemical systems further intensifies the challenge of developing accurate mathematical models for traditional model-based fault diagnosis methods. 
In this context, data-driven fault diagnosis approaches have emerged as a viable and effective alternative. 

Chemical systems are typically equipped with multiple sensors to collect multidimensional operational data. 
However, chemical processes experience reaction delays characterized by time lags between inputs and outputs. As a result, the sampling frequency of sensors is considerably lower compared to that of mechanical systems. 
Commonly measured parameters in chemical processes include temperature, pressure, flow rate, stirring speed, etc. 
Anomalies in any of these parameters can lead to system faults, resulting in a much broader range of potential fault types compared to mechanical systems. 
Consequently, accurately identifying the fault location has become a critical challenge that demands immediate attention.

In recent years, the problem of monitoring multiple operating conditions in chemical systems has attracted considerable attention \cite{song2019multisubspace,zhang2022self,guo2020unsupervised}. 
However, most existing studies have been limited to fault detection and have faced challenges in achieving fault isolation \cite{cong2020anomaly,wang2019data,peng2017multimode,haghani2014quality,chen2020just}. 
This limitation implies that, while the presence of a fault can be identified, the exact fault source cannot be located, thereby significantly increasing the difficulty of fault troubleshooting and maintenance costs. 
Based on current literature, research on MCFD in chemical processes remains relatively limited compared to that on mechanical systems, highlighting substantial potential for future development. 
This presents a valuable opportunity and direction for advancing MCFD research in the context of chemical systems.

\section{Conclusion and prospects}
In this paper, we have provided a comprehensive review of the literature on multi-condition fault diagnosis.  
To ensure clarity, we have mathematically defined the multi-condition scenario.
Subsequently, We have presented a detailed summary of relevant studies and categorized the various MCFD methods. 
We have also discussed the various evaluation metrics employed to assess MCFD methods.  
Additionally, we have analyzed the application scenarios and challenges associated with MCFD methods.
In recent years, multi-condition fault diagnosis has garnered significant attention as an expanding research field.  
Based on a survey of the existing literature, we have summarized the potential future development of MCFD as follows:

\begin{enumerate}
	\item In the existing literature, it is typically assumed that the number of faulty and healthy samples is balanced.
	However, in the actual application scenario, after the system fails, engineers will stop the operation of the equipment as soon as possible to prevent further fault progression and mitigate the risk of more severe consequences.
	As a result, the number of fault samples collected is much lower than that of healthy samples, leading to a bias in the trained model. 
	Therefore, addressing the class imbalance in the design process is of critical importance, particularly for data-driven approaches.
	
	\item With the development of deep learning technology, many DNN-based MCFD methods have demonstrated outstanding performance. 
	However, the black-box nature of deep learning undermines engineers' trust in diagnostic results. 
	Therefore, increasing the interpretability of the model is critical, as this will help with the actual landing and application of MCFD. 
	In this context, \emph{explainable artificial intelligence} (XAI) is considered a promising solution. 
	This approach can assist the model in fault localization and isolation, thereby further reducing operation and maintenance costs.

	\item During the literature review process, it was observed that the majority of existing MCFD papers utilize domain adaptation-based methods. 
	These approaches are generally based on the assumption that newly arrived samples can be accurately subjected to condition identification. 
	However, in actual industrial scenarios, real-time condition identification during online operations is highly challenging due to factors such as environmental noise. 
	Therefore, developing an effective condition identification model is of paramount importance for MCFD.

	\item Most of the algorithms proposed in the existing literature train models offline and are directly deployed at the online stage. 
	Due to the complexity of real-world environments, it is impossible to collect a sufficiently comprehensive dataset to address all possible situations.
	Therefore, updating the model in real-time during the online process becomes a research direction worth exploring. 
	By utilizing new data collected online, the model can be adjusted in real-time, allowing it to adapt to dynamic environments.

	\item Most existing MCFD studies operate under the default assumption that the label distribution remains consistent across different operating conditions while the data distribution varies. 
	However, in real-world scenarios, label shifts may exist between different operating conditions. 
	In such cases, fault diagnosis methods that rely solely on extracting domain-invariant features are unlikely to perform satisfactorily. 
	Therefore, investigating and designing strategies capable of handling label shifts can effectively expand the application scenarios of multi-condition fault diagnosis. 
	
	\item Existing evaluation metrics in the field of MCFD primarily rely on common machine learning metrics. 
	However, to address issues related to data distribution, adopting a probabilistic framework would provide a more scientifically robust approach. 
	From data distribution representation to operating condition identification, fault diagnosis, and evaluation, the development of a systematic probabilistic framework presents a promising and valuable research direction.
	
\end{enumerate}

\bibliographystyle{ieeetr}
\bibliography{MCFD_241119}

\begin{IEEEbiography}
	[{\includegraphics[width=1in,height=1.25in,clip,keepaspectratio]{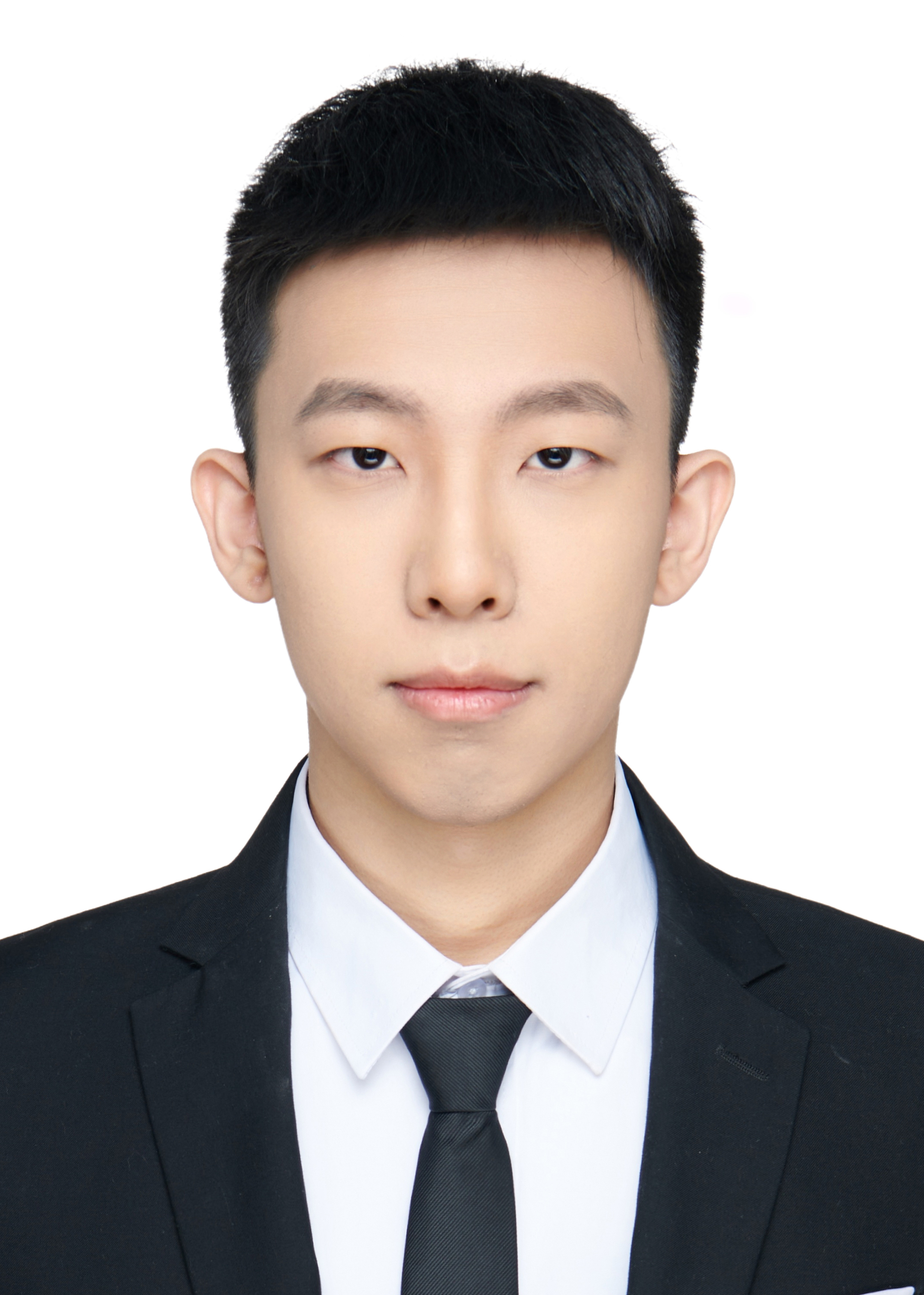}}]{Pengyu Han} 
	received the B.E. degree from the School of Automation, Beijing Institute of Technology, Beijing, China. He is currently pursuing the Ph.D. degree at the Department of Automation, Tsinghua University. His research interests include machine learning and fault diagnosis based on data-driven methods.
\end{IEEEbiography}
\begin{IEEEbiography}
	[{\includegraphics[width=1in,height=1.25in,clip,keepaspectratio]{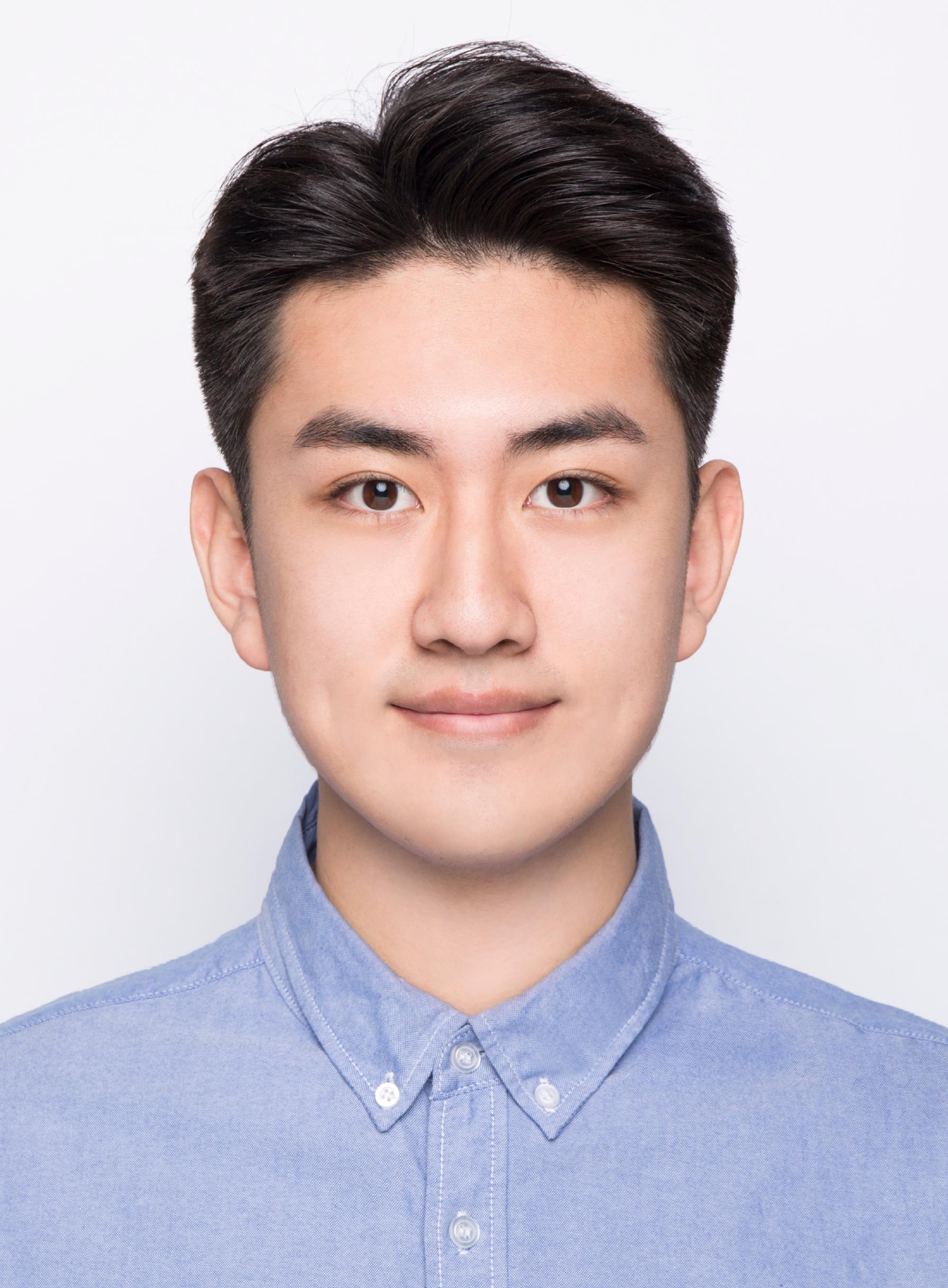}}]{Zeyi Liu}
	received the B.E. degree with the School of Computer and Information Science, Southwest University, China. He is currently pursuing the Ph.D. degree with the Department of Automation, Tsinghua University. His research interests include machine learning, information fusion, fuzzy logic and their applications such as safety assessment and fault diagnosis. He has authored more than 10 papers in refereed international journals and conferences. He severs as a reviewer of refereed journals and conferences such as  \emph{IEEE Trans. Pattern Anal. Mach. Intell., IEEE Trans. Neural Netw. Learn. Syst.}, and \emph{IEEE T. Cybern.}.
\end{IEEEbiography}
\begin{IEEEbiography}
	[{\includegraphics[width=1in,height=1.25in,clip,keepaspectratio]{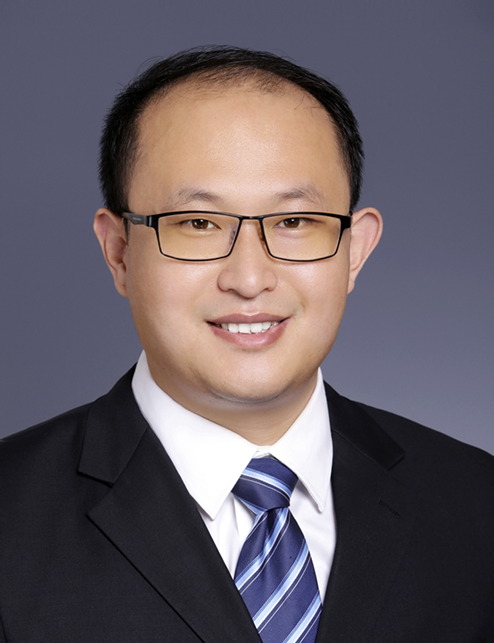}}]{Xiao He}
	(M$'$12--SM$'$20) received the B.Eng. in information technology from the Beijing Institute of Technology, Beijing, China, in 2004, and the Ph.D. degree in control science and engineering from the Tsinghua University, Beijing, China, in 2010.
	
	He is currently a tenured professor with the Department of Automation, Tsinghua University. He has authored more than 100 papers in refereed international journals. His research interests include fault diagnosis and fault-tolerant control, networked systems, cyber-physical systems, as well as their applications. Dr. He is now a Full Member of Sigma Xi, the Scientific Research Society, a Senior Member of the Chinese Association of Automation, and a Senior Member of the IEEE. He is an Associate Editor of the \emph{Control Engineering Practice}.
\end{IEEEbiography}
\begin{IEEEbiography}
	[{\includegraphics[width=1in,height=1.25in,clip,keepaspectratio]{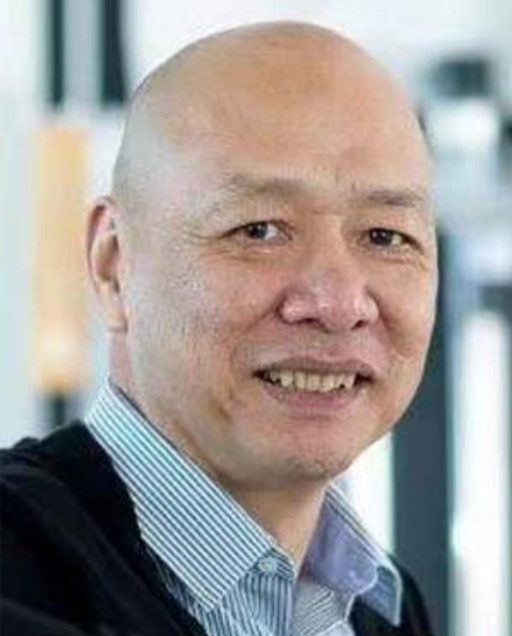}}]{Steven X. Ding}
	received the Ph.D. degree in electrical engineering from the Gerhard-Mercator University of Duisburg, Duisburg, Germany, in 1992. 
	
	He is currently a Full Professor of control engineering and the Head of the Institute for Automatic Control and Complex Systems, the University of Duisburg-Essen, Duisburg. His research interests include model-based and data-driven fault diagnosis, fault-tolerant systems, real-time control, and their application in industry with a focus on automotive systems and chemical processes.
\end{IEEEbiography}
\begin{IEEEbiography}
	[{\includegraphics[width=1in,height=1.25in,clip,keepaspectratio]{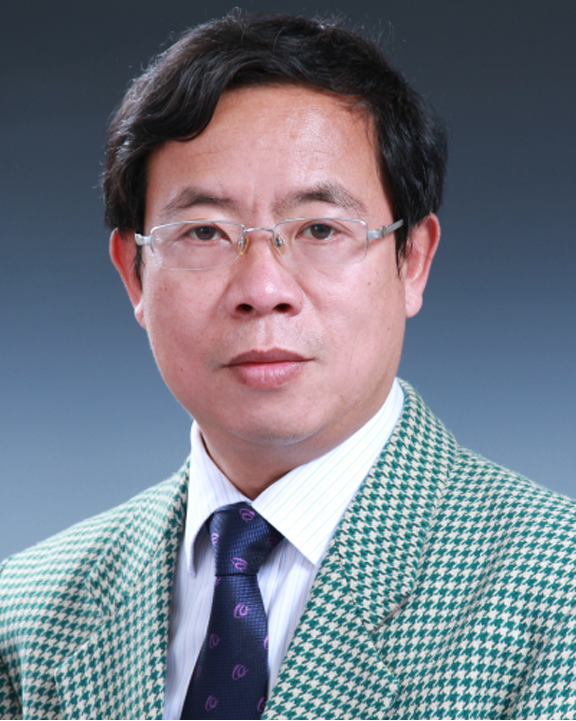}}]{Donghua Zhou}
	(Fellow, IEEE) received the B.Eng., M.Sci., and Ph.D. degrees in electrical engineering from Shanghai Jiao Tong University, Shanghai, China, in 1985, 1988, and 1990, respectively. 
	
	He was an Alexander von Humboldt Research Fellow with the University of Duisburg-Essen, Duisburg, Germany, from 1995 to 1996, and a Visiting Scholar with Yale University, New Haven, CT, USA, from 2001 to 2002. He joined Tsinghua University, Beijing, China, in 1996, where he was promoted to a Full Professor in 1997 and was the Head of the Department of Automation from 2008 to 2015. He was the Vice-President of Shandong University of Science and Technology, China, from 2015 to 2023. He has authored or co-authored over 320 peer-reviewed international journal articles and nine monographs in the areas of fault diagnosis, fault-tolerant control, and operational safety evaluation. 
	
	Dr. Zhou is a fellow of IET, CAA, and AAIA; a member of IFAC TC on SAFEPROCESS; an Associate Editor of \textit{Journal of Process Control} and \textit{Fundamental Research}; and the Vice-Chairperson of Chinese Association of Automation (CAA). He was also the NOC Chair of the Sixth IFAC Symposium on SAFEPROCESS in 2006.
\end{IEEEbiography}

\end{document}